\documentclass[aps, prl, showpacs, twocolumn, amsfonts, amsmath, amssymb, superscriptaddress, floatfix]{revtex4}

\usepackage{graphicx}
\usepackage{dcolumn}
\usepackage{bm}
\usepackage{epsfig}
\usepackage{hyperref}

\begin{document}

\title{Photon bubbles in ultra-cold matter}

\author{J.T. Mendon\c{c}a}

\email{titomend@ist.utl.pt}

\affiliation{IPFN, Instituto Superior T\'{e}cnico, Av. Rovisco Pais 1, 1049-001 Lisboa, Portugal}

\author{R. Kaiser}

\affiliation{Institut Non Lin\'eaire de Nice, UMR 6618,  1361 Route des Lucioles, F-06560 Valbonne, France}

\begin{abstract}

We show that static and oscillating photon bubbles can be excited by diffused light in the laser cooled matter confined in a magneto-optical trap (MOT). The bubble instability is due to the coupling between the radiation field and the mean field oscillations of the ultra-cold gas, and it can provide a source for low frequency turbulence. We consider a diffusion dominated regime, which can be described by a radiation transport equation, coupled with the mean field equations for the cold atom gas. A perturbative analysis shows the occurrence of two different regimes with either oscillating or purely growing bubbles. This work could also be useful to understand similar processes in astrophysics.

\end{abstract}


\maketitle

Atomic physics has played an important role in the development of fundamental laws of quantum physics starting with the Bohr model for the atomic structure. With the advent of spectroscopy, laboratory based experiments in atomic physics have provided many important insights on astrophysical observations. As novel detection techniques in astrophysics allow for ever more subtle phenomena to be explored, ranging from quantum optics to lasing in astrophysical systems \cite{Letokhov2009}, modern tools in atomic physics can be deployed to understand and predict possible features in astrophysics. The recent study of photon scattering in hot atomic vapors has thus lead to the direct observation of Levy flight of photons \cite{Mercadier2009} induced by Doppler broadening and the possibility to implement dilute vapor based random lasing has been investigated \cite{Guerin2010}.
Focusing on the impact of light on the atomic motion, low frequency oscillations of the size of the MOT, reminiscent of oscillations found in variable stars called Cepheids, have been observed \cite{robin1} and numerical simulations predict turbulent like spatio-temporal structures \cite{robin2}. The features are based on a repulsive force due to the exchange of scattered photons between nearby atoms \cite{sesko}. This repulsive force can be described by an effective atomic charge, which explains the coulomb like expansion of the gas \cite{pruvost}. It can also explain the occurrence of collective processes of the plasma type, implying the existence of hybrid phonons, with a lower cut-off \cite{mend1,hugo}, even though until now, these hybrid phonons have not been directly observed.

Here we propose  a new mechanism associated with the laser cooling process, which can lead to the formation of static and oscillating photon bubbles inside the gas, and eventually provide the source for the predicted turbulent-like spatio-temporal structures\cite{robin2}. Photon bubbles have been considered in astrophysical context \cite{Arons1992, Begelman2006, Yorke2007} where huge photon densities are required to have any significant impact on high energy particles. The possibility of modeling astrophysical situations in laboratory experiments has been discussed in high energy physics \cite{Takabe1999, Bulanov2009}. Due to the low kinetic energy of atoms accessible with the development of laser cooling techniques, radiation pressure effects can now be explored in laboratory based experiments with modest photon densities. Novel instabilities based on the coupling between atomic motion and light propagation come in the range of experimental observation. One class of such instabilities is based on "atom-field" coupling, where the index of refraction of the atomic density affects the propagation of light which couples back to atomic motion\cite{Bonifacio1994, Courteille2003, Saffmann1998, Gauthier2011}. This regime can be explored in clouds of cold atoms with low optical thickness, using e.g. large atom-laser detuning. A second regime is based on "atom-intensity" coupling, where coherent effects and interferences can be neglected. This regime valid for larger optical thickness will be considered in this letter. Note that in contrast to past observation of pattern formation in hot atomic vapors \cite{Grantham1991, Gauthier2002} the instabilities based on the coupling to the atomic motion do not require non linear optical response of the atoms.

Our model is based on the coupled photon and atom density evolution equations. The standard approach in the laser cooling community is to distinguish two terms. The first term is due to the coherent coupling between the incident laser beams and the atoms. This term describes the friction and trapping terms producing the confined cold atom sample. A second term, aimed at describing the collective, many atom case, is based on the attenuation of the incident laser beams, often called the shadow effect \cite{Dalibard1988}, and of multiple scattering between the atoms \cite{sesko}. In MOT experiments with very large number of atoms, the multiple scattering (diffusion) terms dominates over the shadow effect. The combined atom-light interaction contain terms proportional to the atom number density. Therefore, a perturbation in the density will lead to a perturbation in the local photon intensity, which can become unstable. If the instability is uniform and isotropic, a local bubble can be formed, where both the photon intensity and atom density are modified.

Let us consider the most relevant situation for the MOT conditions, where the incident laser cooling beams are deflected by multiple scattering, randomizing the photon propagation direction, and the diffusion effects become dominant. In order to describe this new situation we start from the energy transport equation for the photon field which, in a region where radiation sources are absent, can be generally stated as
\begin{equation}
\frac{\partial}{\partial t} I_\omega + \nabla \cdot {\bf S}_\omega = - \gamma_\omega I_\omega .
\label{1} \end{equation}
Here, $\gamma_\omega$ describes photon absorption at the relevant wavelength by e.g. hyperfine Raman scattering.
Even though the loss can often be neglected in realistic description of experimental situations, we keep it in our model as a finite non zero loss term can induce a finite threshold for the instabilities predicted in this letter. The quantity  $I_\omega$ can be defined as the spectral energy density of the electromagnetic radiation $W (\omega, {\bf k})$, integrated over all the possible directions of propagation. If radiation is made isotropic by multiple scattering, the energy flux is determined by a diffusive process, characterized by (see e.g. \cite{van,ishimaru})
$ {\bf S}_\omega = - D \nabla I_\omega$ and $\gamma_\omega = D k_a^2$,
where $k_a$ is the inverse of the energy absorption length. The diffusion coefficient is determined by $D = l^2 / \tau$, where the photon mean free path is $l = 1 / n_a \sigma_L$, where $n_a$ is the atom number density, and $\sigma_L$ the laser atom scattering cross section. The photon diffusion time $\tau$ can be considered as nearly independent from the atom density, as shown by cold atom experiments \cite{labeyrie}. We therefore get $D \propto n_a^{-2}$.

On the other hand, the atom density $n_a$ can be determined in the mean field approximation by the following fluid equations, which determine the atom density $n_a$ and mean velocity ${\bf v}$,
\begin{equation}
\frac{\partial n_a}{\partial t} + \nabla \cdot (n_a {\bf v}) = 0
, \quad
\frac{\partial {\bf v}}{\partial t} + ({\bf v} \cdot \nabla) {\bf v} = \frac{{\bf F}}{M} - \frac{\nabla P}{n_a M} - \nu {\bf v} ,
\label{2} \end{equation}
where $P$ is the pressure, $\nu$ is the damping rate resulting from the viscosity of the gas, and M is the atom mass. The collective force ${\bf F}$, resulting from the exchange of photons between nearby atoms is determined by a Poisson type of equation, $\nabla \cdot {\bf F} = Q n_a$, where $Q = (\sigma_R - \sigma_L) \sigma_L I_\omega$ defines an effective charge for a single atom in the mean field, and $\sigma_R$ and $\sigma_L$ are the atom radiation scattering and atom laser absorption cross sections \cite{sesko,pruvost}. We notice that the charge parameter $Q$ is proportional to the laser intensity, thus providing a coupling between the atom density $n_a$ and the photon intensity $I_\omega$.
In contrast with the astrophysical models for the dynamics of the gas around neutron stars \cite{Arons1992}, young high-mass stars \cite{Yorke2007} or black hole accretion disks \cite{Begelman2006}, the fluid equations for the atomic distribution in our situation do not depend on magnetic fields, preventing a direct mapping of the predictions for photon bubbles in astrophysics.

The two coupled eqs. (1) and (2) provide the starting point for our analysis.
Using a perturbation analysis, we assume that $I_\omega = I_0 + \tilde I$, $n_a = n_0 + \tilde n$, and treat  ${\bf v}$ and ${\bf F}$ as perturbations. Linearizing these equations with respect to the perturbed quantities, and  assuming that they evolve as$\exp (i {\bf q} \cdot {\bf r} - i \Omega t)$, we arrive at the dispersion relation
\begin{equation}
(- i \Omega + D_0 q^2+ \gamma_0 ) ( \Omega^2 + i \Omega \nu - \omega_s^2) = - \beta (\epsilon + i a) \Omega ,
\label{5} \end{equation}
where we have used $ \epsilon = g \nabla^2 I_0$, $ \quad a = g ({\bf q} \cdot \nabla I_0)$,
and  $\omega_s^2 \equiv \omega_p^2 + u_s^2 q^2$, with the effective plasma frequency $\omega_p = (Q n_0 / M)^{1/2}$, and the sound speed $u_s$ \cite{mend1}. We have also used the unperturbed diffusion coefficient $D_0 = 1/ n_0^2 \sigma_L^2 \tau$, $\gamma_0 = D_0 k_a^2$, and $g = 2 D_0 / n_0$. Finally, we have introduced the coupling coefficient $\beta = \omega_p^2 n_0 / I_0$.

At this point it should be noticed that, in the absence of coupling between the photons and the atomic gas ($\beta = 0$), this dispersion relation would describe two independent modes, a purely decaying photon mode, such that $\Omega = - i (D_0 q^2 + \gamma_0)$, where damping results from both diffusion and absorption, and an oscillating fluid mode, determined by $\Omega = \omega_s - i \nu$, which decays due to viscosity. The dispersive properties of this oscillating mode have already been discussed in detail, for both the fluid and kinetic regimes \cite{mend1}. It is now interesting to consider the modes which result from the coupling between the photons and the atomic gas, as described by the new dispersion equation (\ref{5}), when $\beta \neq 0$.

Let us first examine oscillating perturbations with frequency of order $\omega_s$. Using $\Omega = \omega_s + \delta$, and assuming $|\delta| \ll \omega_s$, we obtain solutions of the form
\begin{equation}
\delta =  - \frac{i}{2} \nu -  \frac{\beta (\epsilon + i a)}{2 \omega_s^2} \left( i + \frac{D_0 q^2}{\omega_s} + \frac{\gamma_0}{\omega_s} \right) .
\label{6b} \end{equation}
This can be considered as the perturbed oscillating fluid mode.
We see that, apart from a small correction in the real part of the frequency, we can have a growth or damping rate, determined by the quantity $\Gamma = \Im (\delta) = \Im (\Omega)$.
We focus on the case where the quantity $a$ can be neglected and the following inequality is verified
\begin{equation}
| {\bf q} \cdot \nabla I_0 | \; \ll \quad  \frac{D_0 \omega_p^2}{I_0 \omega_s} \; q^2 \nabla^2 I_0 .
\label{7} \end{equation}
In this case, instability will occur if the following two conditions are simultaneously satisfied
\begin{equation}
\beta \epsilon \; < 0 \; , \quad \frac{| \beta \epsilon |}{\omega_s^2} > \nu .
\label{8} \end{equation}
The first condition implies that, for $\beta > 0$ (i.e. for $\sigma_R > \sigma_L$) oscillating bubbles can only occur in regions with $\nabla^2 I_0 < 0$. Notice that a positive $\beta$ corresponds to the most natural experimental scenario in a MOT. The sign of $\nabla^2 I_0$ depends on boundary conditions or fluctuations of the unperturbed light intensity (e.g. due to imperfections) and can vary across the cloud. The emergence of instabilities thus depends on local initial conditions.
We note that in configurations where the shadow effect would dominate over the repulsion, with  $\beta < 0$ (i.e. $\sigma_R < \sigma_L$), instabilities can occur in regions of $\nabla^2 I_0 > 0$. On the other hand, the second condition in (\ref{8}) determines the instability threshold, where the growth rate has to compensate for the losses due to the viscosity in the ultra-cold gas. In the absence of viscosity, or for conditions well above the threshold, the growth rate attains its maximum value $\Gamma_{max} = | \beta \epsilon | / 2 \omega_s^2$. This will lead to the formation of photon bubbles, as discussed below.
In the opposite situation where $a$ becomes the dominant term in (\ref{6b}) and the inequality (\ref{7}) is reversed, we still have mode instability if $ {\bf q} \cdot \nabla I_0 < 0$, but isotropy is lost and the growth rate will maximize along the gradient of the photon intensity.

\begin{figure}
\includegraphics[angle=0,scale=0.3]{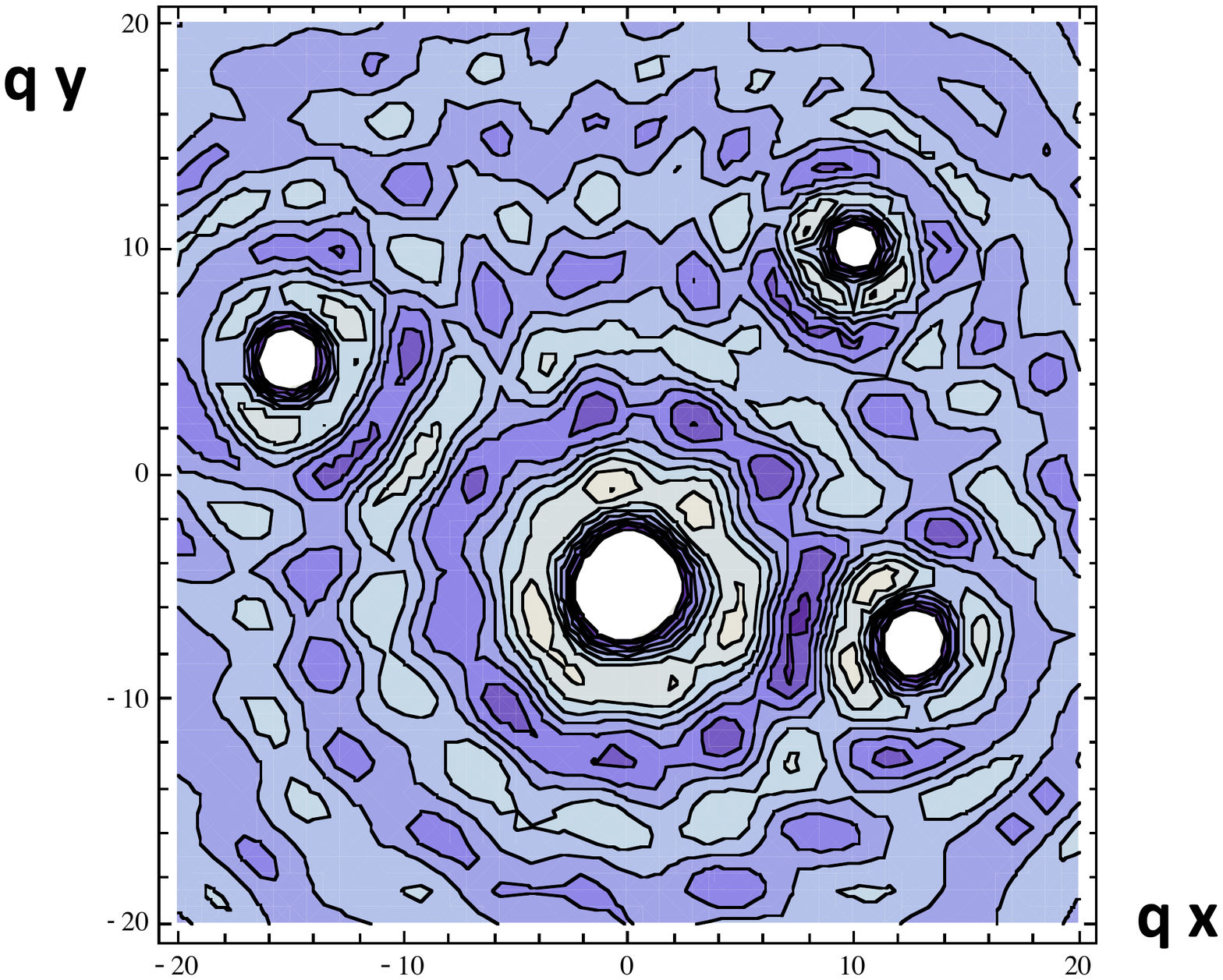}
 \caption{{\sl \label{fig1} Schematic representation of spherical bubbles resulting from the unstable coupling of laser light with the atom mean field. Intensity plots of the function $F (r)$, as determined by eqs. (\ref{10}).}}
\end{figure}

Let us now turn to purely damped (or growing) perturbations, with $\Re (\Omega) \simeq 0$. This can now be considered as the perturbed photon decay mode. Assuming now that $| \Omega|^2 \ll \omega_s^2$, equation (\ref{5}) leads to
\begin{equation}
\Gamma =  \left[ \frac{\beta \epsilon}{\omega_s^2}  - (D_0 q^2 + \gamma_0)  \right]
\label{9} \end{equation}
This shows that unstable modes can indeed exist, for $\beta \epsilon  > 0$.   It should also be noticed that these modes are not exactly purely growing modes, because a residual value of the mode frequency still exists, $\Re (\Omega) \simeq (a D_0 / 2 I_0)$. This quantity goes to zero with $a \equiv  {\bf q} \cdot \nabla I_0$, and therefore vanishes in the spherically symmetric case. For commodity, we will keep calling these modes, purely growing modes.

It is useful to compare these modes with the above oscillating modes. First, they occur for opposite signs of the quantity $\beta \epsilon$, the oscillating modes for a negative sign, and the purely growing modes for a positive sign. The maximum growth rate for the purely growing modes are two times larger than for the oscillating modes, $\Gamma_{max} = \beta \epsilon /  \omega_s^2$. On the other hand, the threshold conditions are different:
i) For the oscillating modes, eq. (\ref{8}) shows that the instability is limited by the viscosity of the atomic gas.
ii) In contrast, for purely growing modes, eq. (\ref{9}) shows that the instability is limited by photon diffusion and losses.

In this comparison we have assumed a negligible $a$, when the instabilities can be seen as isotropic. In this situation, we can associate such unstable modes with the formation of photon bubbles. In order to describe more explicitly the formation of such bubbles, we replace the plane wave modes by spherically symmetric perturbations described by $( \tilde I, \; \tilde n)  \propto F (r) \exp (- i \Omega t)$, where $F (r)$ satisfies the equation
\begin{equation}
\nabla^2 F (r) \equiv \frac{1}{r^2} \frac{\partial}{\partial r} \left( r^2 \frac{\partial}{\partial r} \right) F (r) = - q^2 F (r) ,
\label{10} \end{equation}
where $q^2$ is a positive real quantity. Going back to the perturbed equations, but ignoring the quantity $a \equiv {\bf q} \cdot \nabla I_0$ which would brake the spherical symmetry, we arrive again at the dispersion relation (\ref{5}), but with $ a = 0$. Therefore, our previous analysis for the oscillating and purely growing modes remains valid for such spherically symmetric perturbations.
Equation (\ref{10}) can be written as a spherical Bessel function, with a well known non-singular solution, given by
\begin{equation}
F (r) = \sqrt{\frac{\pi}{2 q r}} J_{1/2} ( q r) = \frac{\sin (q r)}{q r} ,
\label{10b} \end{equation}
where $J_{1/2} (q r)$ is a Bessel function of the first kind. A linear superposition of four such solutions with arbitrary amplitudes and random positions is represented in Fig. \ref{fig1}, for illustration. Of course, superposition brakes down in the nonlinear regime, where instability saturation will occur. For purely growing modes, the saturation level can be easily estimated by replacing in the threshold condition (\ref{9}), the equilibrium diffusion coefficient $D_0$ by its perturbed value $D_0 n_0^2 / (n_0 - |\tilde n|)^2$. Saturation will then occur for
\begin{equation}
| \tilde n |_{sat} \simeq n_0 \left[ 1 - \left( \frac{D_0 q^2}{\Gamma_{max} - \gamma_0} \right)^{1/2} \right]
\label{10bb} \end{equation}
Saturation can then be understood as a result of an increase of the diffusion losses due to a local density depletion.
According to the coupled equations, a local decrease in the atomic number density $\tilde n$ also leads to an increase of the local photon intensity. Near the instability saturation, $\Im ({\Omega}) \sim 0$, for the static bubbles we have the following relation
$\tilde I = - \tilde n /(D_0 q^2 + \gamma_0 )$.
Therefore, we have opposite signs for the perturbations $\tilde I$ and $\tilde n$, as expected.

Finally, it is useful to write the threshold conditions and growth rates in a more explicit form, in terms of the typical time and scale lengths. Introducing the photon intensity scale length $ L$, such that $L^{-2} \equiv I_0^{-1} | \nabla^2 I_0 |$, we can write the threshold and growth rate for the oscillating bubbles as
\begin{equation}
L < \frac{\omega_p}{\omega_s} \frac{2 l}{\sqrt{ 2 \tau \nu}}
\; , \quad \Gamma_{max} = \frac{1}{ \tau} \frac{l^2}{L^2} .
\label{11} \end{equation}
A similar analysis can be made for purely growing bubbles, leading to
\begin{equation}
L <  \frac{\omega_p}{\omega_s} \frac{\sqrt{2}}{(q^2 + k_a^2)^{1/2}}
\; , \quad \Gamma_{max} = \frac{2}{\tau} \frac{l^2}{L^2} .
\label{12} \end{equation}
Given that $\omega_p \simeq \omega_s$, and assuming $k_a^2 \gg q^2$, we get a threshold condition $L <  1 / k_a$, which can be easily satisfied. On the other hand, taking the experimental conditions of \cite{labeyrie}, we have $l \simeq 300 \mu m$, $D \simeq 0.66 m^2 / s$, corresponding to $\tau \simeq 0.1 \mu s$. Thus, even for large $L \gg l$, the growth rates $\Gamma_{max}$ can stay well above $1 s^{-1}$. This can be satisfied in current experimental conditions, with opposite signs of $\beta \epsilon$, for the oscillating and the purely growing cases. The threshold conditions are represented in Fig. 2. Instability occurs for an intensity scale length $L$ lying below the curve (a) for oscillating bubbles, and below curve (b) for purely growing ones, as determined by eqs. (\ref{11}) and (\ref{12}) respectively.
For large wavelengths, such that $q L \leq 1$, the above results are only approximate, and our local analysis should be replaced by a global mode analysis.
\begin{figure}
\includegraphics[angle=0,scale=0.3]{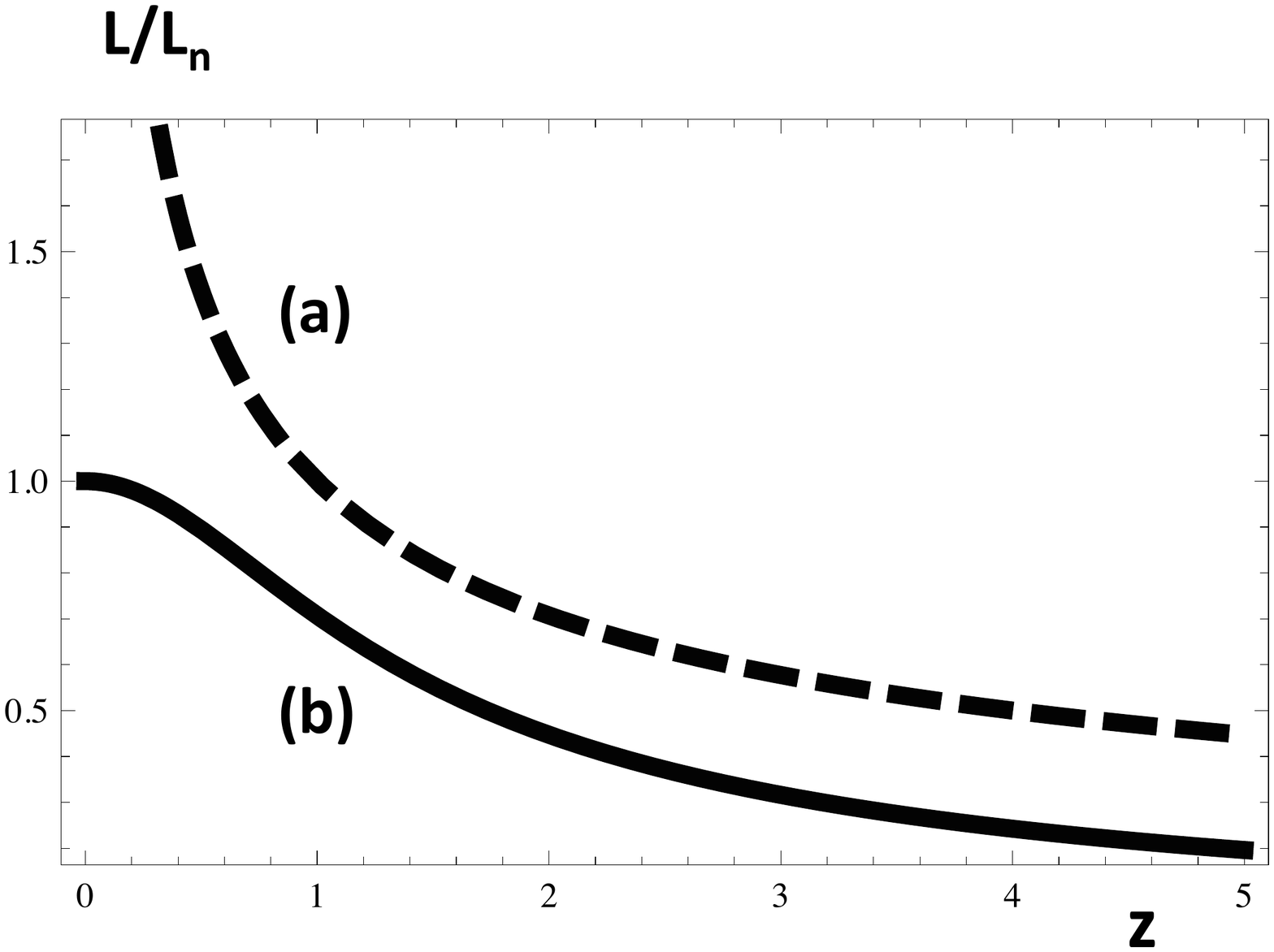}
\caption{{\sl \label{ fig2}Threshold curves for photon bubble formation in the ultra-cold gas: (a) Oscillating bubbles, photon characteristic scale $L / L_n$, versus normalized wavenumber $z = q / k_a$ with $L_n =(\omega_p / \omega_s)  \sqrt{2} / k_a$; (b) Purely growing bubbles,  $L/L_n$ versus $z = \sqrt{\tau \nu}$, with normalization factor $L_n =(\omega_p / \omega_s) l \sqrt{2}$. Instability occurs below theses curves.}}
\end{figure}

In conclusion, we have studied the stability of low frequency perturbations in the photon intensity and atom number density in the laser cooled gas. We have considered the diffusive regime, where multiple scattering of photons by the cold atoms randomizes the direction of photon propagation. Similar to astrophysical systems, radiation pressure is also expected to be important to create turbulence in cold atomic vapors by destabilizing the atomic density distribution with subsequent feedback on radiation transport.
We have used the photon transport equation, coupled to the fluid equations for the ultra-cold gas. Such a coupling is mediated by the atom effective charge, which depends on the photon intensity, and by the diffusion coefficient, which is inversely proportional to the square of atom number density.

We have shown that, both static (or purely growing) and dynamic (or oscillating) photon bubbles can be excited. These two distinct modes result from a perturbation of the two natural modes of the two fluids, the photon gas and the atomic gas, when considered separately. Due to coupling between the two interacting fluids, mediated by both the diffusion coefficient and by the effective charge, the two natural modes eventually become unstable, leading to the formation of purely growing and oscillating structures. In certain experimental conditions, such that the quantity $a \equiv {\bf q} \cdot \nabla I_0 \sim 0$, these unstable structures acquire a spherical symmetry. In contrast, when $a \neq 0$, new oscillating modes appear, which depend on the direction of propagation and which grow faster along the gradient of the photon intensity.
We have therefore characterized the conditions under which photon bubbles can be excited. Nonlinear saturation levels have also been estimated. As shown above, photon bubbles represent a local increase of the photon number density, associated with a local decrease of the atom number density.

The static and the oscillating bubbles occur for different signs of the quantity $\beta \epsilon$, which is typically determined by the sign of $\nabla^2 I_0$.  Such bubble instabilities are isotropic, and a simple spherically symmetric solution was identified, although the basic ingredients of this intensity-atom coupling induced instabilities should also occur in different geometries. These instabilities could provide the driving mechanism for random structures and the excitation of low frequency turbulence in current MOT experiments. Experimental investigations of these structures are in reach with existing fast imaging techniques and we expect various geometries and excitation techniques to provide experimental insight for these instabilities. Finally, it should be added that radiation pressure is also expected to be at the origin of turbulence in astrophysics, opening the way to simulate complex astrophysical situations in the laboratory.


\begin{thebibliography}{10}


\bibitem{Letokhov2009}
V. Letokhov, S. Johansson, " Astrophysical Lasers ", Oxford Uni. Press (2009).

\bibitem{Mercadier2009}
N. Mercadier, W. Guerin, M. Chevrollier and R. Kaiser, Nature Physics  {\bf 5}, 602 (2009),

\bibitem{Guerin2010}
W. Guerin, N. Mercadier, F. Michaud, D. Brivio, L. S. Froufe-Perez, R. Carminati, V. Eremeev, A. Goetschy, S. E. Skipetrov, R. Kaiser, J. Opt ,{\bf 12}, 024002 (2010)

\bibitem{robin1}
G. Labeyrie, F. Michaud and R. Kaiser, {\sl Phys. Rev. Lett.}, {\bf 96}, 023003 (2006).

\bibitem{robin2}
T. Pohl, G. Labeyrie and R. Kaiser, {\sl Phys. Rev. A}, {\bf 74}, 023409 (2006).

\bibitem{sesko}
D.W. Sesko, T.G. Walker and C.E. Wieman,  {\sl J. Opt. Spc. Am. B}, {\bf 8}, 946 (1991).

\bibitem{pruvost}
L. Pruvost, I. Serre, H.T. Duong and J. Jortner, {\sl Phys. Rev. A}, {\bf 61}, 053408 (2000).

\bibitem{mend1}
J.T. Mendon\c ca, R. Kaiser, H. Ter\c cas and J. Loureiro, {\sl Phys. Rev. A}, {\bf 78}, 013408 (2008);
J.T. Mendon\c{c}a, {\sl Phys. Rev. A}, {\bf 81}, 023421 (2010).

\bibitem{hugo}
H. Ter\c cas, J.T. Mendon\c ca and R. Kaiser, {\sl Europhys. Lett.}, {\bf 89}, 53001 (2010).

\bibitem{Arons1992}
J. Arons, Astrophys. J., {\bf 388}, 561 (1992).

\bibitem{Begelman2006}
M. Begelman, Astrophys. J., {\bf 643}, 1065 (2006).

\bibitem{Yorke2007}
N. Turner, E. Quataert, H. Yorke, Astrophys. J., {\bf 662}, 1052 (2007).

\bibitem{Takabe1999}
B. Remington, D. Arnett, R. Drake, H. Takabe, {\sl Science}, {\bf 284}, 1488 (1999).

\bibitem{Bulanov2009}
S. Bulanov, T. Esirkepov, D. Habs, F. Pegoraro, T. Tajima, {\sl Eur. Phys. J. D} {\bf55}, 483–507 (2009).

\bibitem{Bonifacio1994}
R. Bonifacio, L. De Salvo, L. M. Narducci and E. J. D'Angelo, {\sl Phys. Rev. A}, {\bf 50}, 1716 (1994).

\bibitem{Courteille2003}
D. Kruse, C. von Cube, C. Zimmermann, and Ph. W. Courteille , {\sl Phys. Rev. Lett.}, {\bf 91}, 183601 (2003).

\bibitem{Saffmann1998}
M. Saffman, {\sl Phys. Rev. Lett.}, {\bf 81}, 65 (1998).

\bibitem{Gauthier2011}
J.A. Greenberg, B.L. Schmittberger, and D.J. Gauthier, Opt. Express 19, 22535 (2011).

\bibitem{Grantham1991}
J. W. Grantham, H. M. Gibbs, G. Khitrova, J. F. Valley, Xu Jiajin, {\sl Phys. Rev. Lett.}, {\bf 66}, 1422 (1991).

\bibitem{Gauthier2002}
R. S. Bennink, V. Wong, A. M. Marino, D. L. Aronstein, R. W. Boyd, C. R. Stroud, S. Lukishova, D. J. Gauthier, {\sl Phys. Rev. Lett.}, {\bf 88}, 113901 (2002).

\bibitem{Dalibard1988}
J. Dalibard, {\sl  Opt. Commun.}, {\bf 68}, 203 (1988).
\bibitem{van}
M.C.W. van Rossum and Th.M. Nieuwenhuizen, {\sl Rev. Mod. Phys.}, {\bf 71}, 313 (1999).

\bibitem{ishimaru}
A. Ishimaru, {\sl Wave propagation and scattering in random media}, Academic Press, New York (1978), chap.9.

\bibitem{labeyrie}
G. Labeyrie, E Vaujour, C.A. M\"uller, D. Delande, C. Miniatura, D. Wilkowski and R. Kaiser,
{\sl Phys. Rev. Lett.}, {\bf 91}, 223904 (2003).


\end{thebibliography}
\end{document}